\documentclass[fleqn,10pt]{wlscirep}
\usepackage{amsfonts,amsmath,bm}
\usepackage[utf8]{inputenc}

\newcommand{\rr}{\bm{r}}
\newcommand{\kk}{\bm{k}}

\newcommand{\kp}{\mbox{$\bm{k}\!\cdot\!\bm{p}$ }}

\newcommand{\cp}{\mathrm{c.p.}}
\newcommand{\hc}{\mathrm{h.c.}}
\newcommand{\es}{\eta}

\newcommand{\Eg}{E_{\mathrm{g}}}
\newcommand{\Eso}{E_{\mathrm{g}} +\Delta_{\mathrm{SO}}}
\newcommand{\tcal}{\mathcal{T}}
\newcommand{\ttcal}{\widetilde{\mathcal{T}}}
\newcommand{\hS}{\hat{S}}

\usepackage{graphicx}

\DeclareMathOperator{\tr}{Tr}

\title{Limited accuracy of conduction band effective mass equations for semiconductor quantum dots}  
\author[1]{Adam Mielnik-Pyszczorski}
\author[1]{Krzysztof Gawarecki}
\author[1*]{Pawe{\l} Machnikowski}
\affil[1]{Department of Theoretical Physics, Faculty of Fundamental
  Problems of Technology, Wroc{\l}aw University of Science and
Technology, 50-370 Wroc{\l}aw, Poland}
\affil[*]{Pawel.Machnikowski@pwr.edu.pl}  

\begin{abstract}
Effective mass equations are the simplest models of carrier states in a semiconductor
structures that reduce the complexity of a solid-state system to Schr\"odinger- or
Pauli-like equations resempling those well known from quantum mechanics textbooks.
Here we present a systematic derivation of a conduction-band effective mass equation for a
self-assembled semiconductor quantum dot in a magnetic field from the 8-band \kp
theory. The derivation allows
us to classify various forms of the effective mass equations in terms of a hierarchy of
approximations. We assess the accuracy of the approximations in calculating selected
spectral and spin-related characteristics. We indicate the importance of 
preserving the off-diagonal terms of the valence band Hamiltonian and argue that an
effective mass theory cannot reach satisfactory accuracy without self-consistently
including non-parabolicity corrections and renormalization of \kp parameters. Quantitative
comparison with the 8-band \kp results supports the phenomenological Roth-Lax-Zwerdling 
formula for the $g$-factor in a nanostructure.
\end{abstract}

\begin{document}

\flushbottom
\maketitle

\section*{Introduction} 
\label{sec:intro}

The \kp method is a well-established approach to calculating the electronic and magnetic
properties of 
bulk semiconductors \cite{yu05}. It has also been applied to semiconductor nanostructures
within the envelope function approximation \cite{Willatzen2009,winkler03}, in which a
carrier state is described as a superposition of contributions from different bands,
with local amplitudes smoothly varying in space and referred to as the \textit{envelope
wave  functions}.  
Its accuracy is in many cases sufficient  in many applications throughout
mesoscopic physics, even when subtle details of the spectrum are
considered \cite{Ardelt2016b,Kadantsev2010,doty09}. At the same time, it offers much lower
computiational costs, better scalability to larger systems and much higher transparency
than the more exact atomistic approaches 
\cite{zielinski10,usman11,narvaez05}.
The applicability of the \kp method for modulated systems (nanostructures) has been confirmed by
providing a rigorous derivation from the full Schr\"odinger equation
\cite{burt92,Foreman1993}. Recent developments bring the \kp methods down to the atomistic
level\cite{pryor15}.  Currently, the 8-band \kp model  is a widely
tested and generally 
trusted standard for calculating carrier states (including magnetic effects) in mesoscopic
structures, in particular in strained, self-assembled
systems\cite{andlauer08,eissfeller11,voon09k,ehrhardt14}.

In many cases, when the focus is on the electron states, a single-band description is
desired, which would offer further reduction of 
computational costs, make it possible to approximate the problem by a simple, analytically
solvable model, and provide still more straight-forward interpretation referring to the
well-known properties of the usual Schr\"odinger or Pauli equations. Such models, known
as \textit{effective mass} equations, have been used for decades to describe smoothly
varying perturbations of semiconductor systems, like shallow impurities
\cite{Luttinger1955} and later to semiconductor heterostructures
\cite{Dingle1974,Maan1982}.  In the case of quantum dots (QDs), an example  
of a simple approach based on the effective mass approximation is the ``particle in a box
model'' \cite{wojs96} that can further be approximated by the celebrated (and analytically
solvable) Fock-Darwin model \cite{fock28,darwin30}, which once offered general
understanding of QD properties \cite{jacak98a} and is still widely used, at least as the first
approximation to many problems.

In a periodic system, the reduction of the multi-band \kp model to an effective mass
theory is achieved via a quasi-degenerate perturbation theory that
has its roots in quantum chemistry and is referred to as L\"owdin partitioning
\cite{lowdin51}. By means of this procedure, one gets for the conduction band (cb) in the lowest
(second) order a 
Schr\"odinger- (or Pauli-) like equation for the envelope function with a constant
effective mass and Land\'e factor expressed via well-known formulas in terms of the
band-edge energies. 
Two problems are encountered in an attempt to heuristically generalize such an equation to a
modulated system (a nanostructure): First, it is generally accepted that (in analogy to
the multi-band Hamiltonian), the components of the wave vector $\kk$ should be replaced by
derivatives (momentum operators in the coordinate representation). However, at the same
time, the band edges and other parameters become position-dependent via their 
dependence on local composition and strain. This leads to a Schr\"odinger-like equation
with a
position-dependent effective mass and Land\'e factor and the ordering between the momenta
and these two parameters starts to play a role. This ordering is arbitrary if the
effective mass model is introduced as a heuristic extension of the bulk
formulas. Arguments based on solvable models of abrupt interfaces
\cite{Einevoll1988,Thomsen1989} lead to the conclusion 
that the only physically correct ordering in a non-homogenous medium is the most common
``$k(1/m^{*})k$''. This conclusion has been formally verified by rigorous derivation from
the exact multi-band envelope function equations \cite{burt92,Willatzen2009}. 

In this work we would like to address a more practical question of the 
achievable quantitative accuracy of the cb effective mass theories for
quantum dot (QD) systems.
Thus, the goal of this paper is to rigorously derive a family of effective mass Hamiltonians for a
cb electron in a strained, inhomogeneous system by applying a systematic
series of approximations to the L\"owdin partitioning of the 8-band \kp
Hamiltonian in the envelope function approximation and to validate their predictions against
the results of the 8-band model with respect to the energy spectrum and Land\'e factors. 
By \textit{effective mass Hamiltonians} we understand cb Hamiltonians for a
carrier in an external magnetic field, obtained by a unitary transformation (partitioning) that
eliminates the coupling 
between the conduction and valence bands of the original Hamiltonian, 
that contain quadratic (in momentum $\kk$) kinetic energy terms and may contain cubic
spin-orbit 
couplings. We will show that reproducing the results from the 8-band theory by an effective
mass equation is possible with limited precision and the form of the equation required to
achieve the best available accuracy is much more complicated than the usual Schr\"odinger- or
Pauli-like equation for the envelope function. On the other hand, the Land\'e $g$ factor
of the nanostructure can be reasonably estimated by the simple semi-phenomenological
Roth-Lax-Zwerdling formula. 


\section*{The starting point: 8-band \mbox{$\bm{k}\!\cdot\!\bm{p}$} model} 

The 8-band \kp Hamiltonian in the envelope function approximation is defined in the block
notation as\cite{winkler03} 
\begin{equation}\label{H-block}
H=\left(\begin{array}{ccc}
H_{\mathrm{6c6c}} & H_{\mathrm{6c8v}} & H_{\mathrm{6c7v}} \\
H_{\mathrm{8v6c}} & H_{\mathrm{8v8v}} & H_{\mathrm{8v7v}} \\
H_{\mathrm{7v6c}} & H_{\mathrm{7v8v}} & H_{\mathrm{7v7v}} \\
\end{array}\right),
\end{equation}
where the blocks refer in the standard way to the cb
(6c), the $j=3/2$ valence band (vb, 8v) and the $j=1/2$ (spin-orbit
split-off) vb (7v) and are explicitly given by\cite{winkler03,eissfeller11} 
\begin{subequations}
\begin{align}
H_{\mathrm{6c6c}}  
                         & = E_{\mathrm{c}} +V_{\mathrm{p}}            +a_{\mathrm{c}}\tr\es
                  + \frac{\hbar^{2}}{2m_{0}}  \left( k_{x}  A'_{c} k_{x}
                    +  \frac{i}{2}  k_{[x}g' k_{y]} \sigma_{z} + \cp \right) 
  \label{6c6c}\\
H_{\mathrm{8v8v}}  
                         & = E_{\mathrm{v}} -\frac{\hbar^{2}}{2m_{0}} \left \{
                          k_{x} \gamma'_{1}  k_{x} - 2 
                            \left (  J^{2}_{x}  - \frac{1}{3} J^{2}  \right )
                          k_{x} \gamma'_{2} k_{x} 
                        -   \{J_{x},J_{y}\} k_{\{x} \gamma'_{3} k_{y\}}    
                          + \cp   \right\}+ \frac{1}{2\sqrt{3}}  \left[ \{ J_{x},J^{2}_{y} -
                          J^{2}_{z}  \}  \{ C_{k},k_{x} \}   + \cp  \right] \nonumber\\
                          &\quad +a_{\mathrm{v}} \tr{\es} - b_{\mathrm{v}} \left[ 
                          \left(   J^{2}_{x}  - \frac{1}{3} J^{2} \right)\es_{xx} + \cp
                            \right]
                            - \frac{d_{\mathrm{v}}}{\sqrt{3}} \left[ \{J_{x},J_{y}\}
                          \eta_{xy}  + \cp \right] -i \frac{\hbar^{2}}{m_{0}} \left[ k_{[x} \kappa' k_{y]}
                         J_{z} + k_{[x} q k_{y]} J^{3}_{z} + \cp \right ],
  \label{8v8v}\\
H_{\mathrm{7v7v}} &= E_{\mathrm{v}} + V_{\mathrm{p}}
                    -\Delta_{0} - \frac{\hbar^{2}}{2m_{0}} \left( k_{x} \gamma'_{1} k_{x}
                    +\cp \right) +
                    a_{\mathrm{v}} \tr{\es} 
                    -i \frac{\hbar^{2}}{m_{0}} \left[   k_{[x} \kappa' k_{y]}
                    \sigma_{z} + \cp \right ] 
                    - \left ( \mu_{B} B_{z} \sigma_{z} + \cp \right ),
  \label{7v7v}\\	
H_{\mathrm{6c8v}} & 
                          = \sqrt{3}\bm{T}\cdot\tilde{\kk} P + i\frac{\sqrt{3}}{2}(T_{x} 
                          k_{\{y} B_{\mathrm{8v}}^{+} k_{z\}} +\cp)             \nonumber \\
  &\quad +\frac{\sqrt{3}}{2}                          \left[ (T_{xx}-T_{yy}) \left (\frac{2}{3} k_{z}
                          B_{\mathrm{8v}}^{-} k_{z} - \frac{1}{3} k_{x} B_{\mathrm{8v}}^{-} k_{x} -
                          \frac{1}{3} k_{y} B_{\mathrm{8v}}^{-} k_{y} \right )
                    -T_{zz}(k_{x} B_{\mathrm{8v}}^{-} k_{x} - k_{y}
                          B_{\mathrm{8v}}^{-} k_{y}) \right],                           \label{6c8v} \\
H_{\mathrm{6c7v}} & 
                          = -\frac{1}{\sqrt{3}}\bm{\sigma}\cdot\tilde{\kk} P 
                          -\frac{i}{2\sqrt{3}}(\sigma_{x}  k_{[y}B_{\mathrm{7v}}
                    k_{z]} +\cp),   
                    \label{6c7v}\\
H_{\mathrm{8v7v}} & = 
                          -\frac{\hbar^{2}}{2m_{0}} \left \{ 
                          -6  (T^{\dagger}_{xx} k_{x} \gamma'_{2} k_{x} + \cp
                          )  -6  (T^{\dagger}_{xy} k_{\{x} \gamma'_{3} k_{y\}} + \cp )
                    \right\}
                    - i \frac{ \sqrt{3}}{2}  \left ( T^{\dagger}_{yz}
                         \{ C_{k},k_{x} \} + \cp \right ) \nonumber\\
                        & \quad -3 b_{\mathrm{v}} \left( T^{\dagger}_{xx} \eta_{xx} +
                          \cp \right) -  \sqrt{3} d_{\mathrm{v}} \left( 2 T^{\dagger}_{xy}
                          \eta_{xy} + \cp \right)   - i \frac{3 \hbar^{2}}{2 m_{0}} 
                           \left[  k_{[x} \kappa' k_{y]} T^{\dagger}_{z} + \cp \right ]
                          - 3 \left ( \mu_{B} B_{z} T^{\dag}_{z} +  \cp \right ).  \label{8v7v}
\end{align}
\end{subequations}
Here $\{\mathcal{O}_{1}, \mathcal{O}_{2}\}
=\mathcal{O}_{1}\mathcal{O}_{2}+\mathcal{O}_{2}\mathcal{O}_{1}$,  
$k_{\{i}\mathcal{O} k_{j\}} =
 k_{i} \mathcal{O} k_{j} + k_{j} \mathcal{O} k_{i}$, $k_{[i} \mathcal{O} k_{j]} = k_{i} \mathcal{O} k_{j}
 - k_{j} \mathcal{O} k_{i}$ for any operators $\mathcal{O}$, $\mathcal{O}_{1}$, $\mathcal{O}_{2}$; 
$E_{\mathrm{c}}$ and $E_{\mathrm{v}}$ are the cb and vb edges, respectively
($E_{0}=E_{\mathrm{c}}-E_{\mathrm{v}}$ is the fundamental band gap in a bulk crystal);
$\es$  is the strain tensor corresponding to the static 
deformation due to the lattice mismatch;
$\kk=-i\nabla+e\bm{A}/\hbar$, where $\bm{A}$ is the vector potential
of the magnetic field $\bm{B}$;
$\tilde{\kk}=\kk (\mathbb{I}-\es)$; 
$V_{\mathrm{p}}$ is the piezoelectric potential;  $m_{0}$ is the free electron mass;
$A'_{c}$, $g'$ and $\kappa'$  are given by\cite{winkler03}
\begin{equation} 
 A'_{c} \equiv \frac{m_{0}}{m'} =  \frac{m_{0}}{m^{*}} - \frac{2}{3} \frac{E_{P}}{E_{0}} -
                \frac{1}{3} \frac{E_{P}}{E_{0}+\Delta_{0}}, \quad
  g'= 2, \quad  \kappa'= -\frac{1}{3} \left ( \gamma'_{1} - 2 \gamma'_{2} - 3 \gamma'_{3} + 2 \right );
\label{g-prim}
\end{equation}
$P=\hbar(E_{P}/2m_{0})^{1/2}$ (see below for the definition of $E_{P}$);
$\gamma_{i}'$ and $\kappa'$ are the Luttinger parameters with removed contributions
from the $\Gamma_{6}$ cb, 
$\gamma'_{1} =  \gamma_{1} -  E_{P}/(3 E_{0} + \Delta_{0})$, 
$\gamma'_{2,3} =  \gamma_{2,3} - E_{P}/(6 E_{0} + 2\Delta_{0})$,
$\mu_{B}$  is the Bohr magneton;
$q$ is another parameter of the Luttinger Hamiltonian;
$B_{7\mathrm{v}}=(P'Q/i)[1/(E_{0}-E_{0}'-\Delta_{0}')-1/(\Delta_{0}+E_{0}'+\Delta_{0}')]$,
$B_{8\mathrm{v}}^{\pm}=(P'Q/2i)[\pm 1/(E_{0}-E_{0}'- \Delta_{0}')\mp 1/(E_{0}'+\Delta_{0}')
+ 1/(E_{0}-E_{0}')-1/E_{0}']$, where $P'$ and $Q$ are couplings to higher conduction
bands; 
$\sigma_{i}$ are Pauli matrices; $J_{i}$ are matrices of the $j=3/2$
representation of angular momentum; $T_{i}$ are matrix representations
of a vector operator between $j=1/2$ and $j=3/2$ 
states, i.e., $T_{x}=(T^{(1)}_{-1}-T^{(1)}_{+1})/\sqrt{2}$, 
$T_{y}=-(T^{(1)}_{-1}+T^{(1)}_{+1})/\sqrt{2}$, $T_{z}=T^{(1)}_{0}$,
with the matrix elements of the spherical components $T^{(1)}_{q}$
given in terms of the Clebsch-Gordan coefficients 
$\langle j_{1}j_{2};m_{1}m_{2}|jm\rangle$ by the 
Wigner-Eckart theorem, 
$\langle m |T^{(1)}_{q} | m'\rangle = -\sqrt{2/3}\langle 3/2,m';1,q|1/2,m\rangle$,
for $m=\pm 1/2$, $m'=-3/2,\ldots,3/2$; and 
$T_{ij} = T_{i}J_{j}+T_{j}J_{i}$.
The system is placed in an axial magnetic field. 
In numerical calculations we use gauge-invariant discretization scheme\cite{andlauer08}
for the covariant derivative.

The material parameters used in our $\kp$ calculations are given in
Table~\ref{tab:param}. In order to avoid $A'_{c} < 0$, which would break the ellipticity condition,
we rescale $E_{P}$ to obtain $A'_{c} = 1$, which gives\cite{birner11} 
$E_{P} = 
        (m_{0}/m^{*} - 1 )  E_{0}(E_{0}+\Delta_{0})/(E_{0}+2\Delta_{0}/3)$.
Due to inconsistency of the reported
values\cite{winkler03,lawaetz71}, we calculate $q$ using the perturbative formula\cite{eissfeller12}
$q = (2/9)E_{Q}[1/E'_{0} -1/(E'_{0}+\Delta'_{0})]$,
where $E_{Q}$, $E'_{0}$ and $\Delta'_{0}$ are $14$ band $\kp$
parameters\cite{winkler03}. We account for the strain within a continuous elasticity
approach\cite{pryor98b}. Piezoelectric field in the system is calculated up to the second
order in the polarization\cite{bester06} with the parameters taken from
Ref.~\cite{caro15}.   

\begin{table}
	\begin{tabular}{llll}
		\hline
		& GaAs & InAs & Interpolation for $\mathrm{In_{x}Ga_{1-x}As}$\\
		\hline
		$E_{\mathrm{v}}$ & $0.0$~eV & $0.21$~eV & linear\\
		$E_{\mathrm{0}}$ & 1.519~eV & 0.417~eV & $0.417x+1.519(1-x)-0.477x(1-x)$\\
		$E'_{\mathrm{0}}$ & 4.488~eV & 4.390~eV & linear\\
		$E_{\mathrm{Q}}$ & 17.535~eV & 18.255~eV & linear\\
		$m^{*}$ & $0.0665m_{0}$ &
                 $0.0229m_{0}$ & $  [ 0.0229x + 0.0665(1-x)-0.0091x(1-x)] m_{0}$ \\
		$\Delta$ & 0.341~eV & 0.39~eV& $0.39x+0.341(1-x)-0.15x(1-x)$ \\ 
		$\Delta'_0$ & 0.171~eV & 0.25~eV& linear \\ 
		$a_{\mathrm{c}}$ & -7.17~eV & -5.08~eV & $-5.08x-7.17(1-x)-2.61x(1-x)$\\
		$a_{\mathrm{v}}$ & 1.16~eV & 1.00~eV & linear\\
		$b_{\mathrm{v}}$ & -2.0~eV & -1.8~eV & linear\\
		$d_{\mathrm{v}}$ & -4.8~eV & -3.6~eV & linear\\ 
		$\gamma_{\mathrm{1}}$ & 6.98 & 20.0 & $1/ \left [ (1-x)/6.98 + x/20.0 \right ]$ \\
		$\gamma_{\mathrm{2}}$ & 2.06 & 8.5 & $1/\left [ (1-x)/8.5 + x/2.06 \right ]$\\
		$\gamma_{\mathrm{3}}$ & 2.93 & 9.2 & $1/\left [ (1-x)/9.2 + x/2.93 \right
                                                     ]$\\  
          $P'$ & $4.780i$ & $0.873i$ & linear \\
          $Q$ & 8.165 & 8.331 & linear \\
		\hline
	\end{tabular} 
	\caption{\label{tab:param}Material parameters used in the calculations.\cite{Vurgaftman2001,winkler03}}
\end{table}

\section*{Derivation of the effective mass Hamiltonian}

The essence of the method \cite{lowdin51,cohen98}
is to perturbatively decouple the group of states of 
interest from all the other states by using a canonical transformation
$T=e^{S}$, with an anti-hermitian operator $S$, in order to obtain a
transformed Hamiltonian $\tilde{H}=THT^{\dag}$, in which the
inter-band terms (treated as a perturbation) are approximately eliminated. 
We will use a modified version of the van Vleck quasi-degenerate
perturbation theory \cite{VanVleck1929} as presented in 
Ref.~\cite{Shavitt1980}. 
The Hamiltonian is split into its block-diagonal  and block-off-diagonal parts (coupling
states within a single group of states and between the two groups of states, respectively),
$H=H^{(\mathrm{d})}+H^{(\mathrm{od})}$.
The operator $S$ is required to
have null matrix elements within the groups (that is, it has to be
block-off-diagonal). The group of states of interest here are the 
cb states. Since we start from the 8-band \kp theory, the other group of
states are vb states.

We define the superoperator $\hS$ representing the adjoint action of $S$ on the algebra of
operators: 
$\hS \mathcal{O} = [\mathcal{O} ,\hS]$, $\hS^{2}\mathcal{O} = [[\mathcal{O},\hS],\hS]$, 
etc. for any operator $\mathcal{O}$. Functions of $\hS$ are
defined via their power series expansion. Then, from the Campbell-Baker-Hausdorff
expansion, 
$ \tilde{H} = e^{S} H e^{-S} = e^{\hS}H = \cosh\hS H + \sinh\hS H$.
Even functions of $\hS$ transform block-diagonal operators into block-diagonal
operators and block-off-diagonal operators into block-off-diagonal
operators, and the opposite holds for odd functions of $\hS$. Therefore, the
block-diagonal and block-off-diagonal parts of $\tilde{H}$ are
\begin{equation}
\tilde{H}^{(\mathrm{d})}  = \cosh\hS H^{(\mathrm{d})} + \sinh\hS H^{(\mathrm{od})},
\quad
\tilde{H}^{(\mathrm{od})} = \sinh\hS H^{(\mathrm{d})} + \cosh\hS H^{(\mathrm{od})}.
\label{Ht} 
\end{equation}
We require that the resulting Hamiltonian is block-diagonal, hence
$\tilde{H}^{(\mathrm{od})}=0$. Inverting Eqs.~\eqref{Ht} then yields
\begin{equation}
H^{(\mathrm{od})}=-\sinh\hS \tilde{H}^{(\mathrm{d})}. 
\label{sinhG}
\end{equation}
In our case $H^{(\mathrm{od})}$ consists of
$H_{\mathrm{6c8v}},H_{\mathrm{6c7v}}$ [Eq.~\eqref{6c8v} and Eq.~\eqref{6c7v}]
and their hermitian conjugates, 
$H^{(\mathrm{od})} = H_{\mathrm{cv}}+\hc =
H_{\mathrm{6c8v}} \oplus H_{\mathrm{6c7v}} +\hc$,
hence it contains terms linear and quadratic in $\bm{k}$. Since $\tilde{H}^{(\mathrm{d})}$
contains $k$-independent terms, $S$ must be $O(k)$.
According to the first of Eqs.~\eqref{Ht}, the leading order terms in $S$ yield quadratic corrections
to $\tilde{H}^{(\mathrm{d})}$. We therefore take into account only the linear term in
Eq.~\eqref{sinhG} and write
\begin{equation}
H^{(\mathrm{od})}=-\hS \tilde{H}^{(\mathrm{d})}=[S,\tilde{H}^{(\mathrm{d})}]. 
\label{S-kom}
\end{equation}
The neglected corrections are $O(k^{3})$ and would lead to quartic terms in
$\tilde{H}^{(\mathrm{d})}$, which is beyond the effective mass approximation, even with
spin-orbit terms. Quantitatively, the linear truncation according to Eq.~\eqref{S-kom}
amounts to neglecting corrections of relative magnitude $\Delta E/E_{0}$, where $\Delta E$
is the interband energy separation due to confinement. Taking $\Delta E$ as the excitation
energy in the direction of strongest confinement ($\sim 200$~meV) one gets a rough
estimate of 20\% for the error due to truncation.

The remaining, block-diagonal part of the transformed Hamiltonian is now written as
$\tilde{H}^{(\mathrm{d})} =H_{0}+\tilde{H}_{\mathrm{c}}'\oplus \tilde{H}_{\mathrm{v}}'$,
where $H_{0}$ is assumed to
be diagonal in a certain basis $|\alpha,i\rangle$ (where $\alpha$ denotes a group of
states and $i$ are individual states within these groups), and is selected in such a way that the
remaining parts are in some sense small. $\tilde{H}_{\mathrm{c}}'$ and
$\tilde{H}_{\mathrm{v}}'$ denote the cb and vb blocks of $\tilde{H}$,
respectively, with the corresponding parts of $H_{0}$ subtracted. 
In the problem at hand, we choose for $H_{0}$ a diagonal Hamiltonian which is constant and
proportional to unity
within each of the 6c, 8v and 7v bands and approximately represents the band edges in a
strained QD. We will denote the respective energy values by $\overline{E}_{\alpha}$,
$\alpha=\mathrm{6c,8v,7v}$. One can understand them as the average band edges
corrected by 
hydrostatic strain (while the splitting between heavy and light
holes within the 8v band is not included).
The operator $S$ is written as $S=S_{\mathrm{cv}}+ \hc$, where
$S_{\mathrm{cv}}$ denotes one of the two off-diagonal blocks (a $2\times 6$ matrix
in the standard \kp matrix notation).
The effective band gaps are denoted by $\overline{E}_{\mathrm{6c}}-\overline{E}_{\mathrm{8v}}=\Eg$ and 
$\overline{E}_{\mathrm{6c}}-\overline{E}_{\mathrm{7v}}=\Eso$.
Finally we define a diagonal operator 
$\hat{\Delta}=\mathrm{diag}(\Eg,\Eg,\Eg,\Eg,\Eso,\Eso)$, 
where the entries correspond to the 6 valence bands of the 8-band \kp model.
Then,  from Eq.~\eqref{S-kom}, one finds
\begin{equation}\label{sylvester}
\tilde{H}_{\mathrm{c}}'S_{\mathrm{cv}} - S_{\mathrm{cv}}(\tilde{H}_{\mathrm{v}}'-\hat{\Delta}) = 
-H_{\mathrm{cv}}.
\end{equation}
Note that the arbitrariness of choosing the diagonal Hamiltonian $H_{0}$ is removed here,
as the subtracted energies (the operator $\hat{\Delta}$) are added back to the remaining
part of the Hamiltonian.

Eq.~\eqref{sylvester} has a structure of a Sylvester equation but the operators appearing here are not
finite-dimensional matrices. One can treat this equation as a matrix one, in the sense of
the block notation over the subbands, but then the problem of non-commutativity of the
matrix elements appears (due to non-commutativity of $\kk$ with position-dependent
quantities), precluding the application of standard algebraic methods for 
solving this equation. In order to overcome this difficulty, we expand the operators
$S_{\mathrm{cv}}$ and $H_{\mathrm{cv}}$ in powers of $\kk$, 
\begin{equation*}
S_{\mathrm{cv}}=X^{(0)}+\sum_{j}k_{j}X^{(1)}_{j}+ \sum_{jl}k_{j}X^{(2)}_{jl}k_{l}, \quad
H_{\mathrm{cv}}= -\sum_{j}k_{j}C^{(1)}_{j}- \sum_{jl}k_{j}C^{(2)}_{jl}k_{l},
\end{equation*} 
where the coefficients $C^{(1)}$ and $C^{(2)}$ are defined by comparison with the explicit
form of Eq.~\eqref{6c8v} and Eq.~\eqref{6c7v}.
Then, upon rearrangement of terms one gets from Eq.~\eqref{sylvester} in the subsequent
(formal) orders in $\kk$ 
\begin{subequations}
\begin{align}
\tilde{H}_{\mathrm{c}}' X^{(2)}_{jl} - X^{(2)}_{jl} (\tilde{H}_{\mathrm{v}}'-\hat{\Delta}) & =
  C^{(2)}_{jl},  \label{Sylv-a} \\
\tilde{H}_{\mathrm{c}}' X^{(1)}_{j} - X^{(1)}_{j} (\tilde{H}_{\mathrm{v}}'-\hat{\Delta}) &=
  C^{(1)}_{j} + [\tilde{H}_{\mathrm{c}}',k_{l}]X^{(2)}_{lj} 
+ X^{(2)}_{jl} [k_{l},\tilde{H}_{\mathrm{v}}'-\hat{\Delta}],  \label{Sylv-b} \\
\tilde{H}_{\mathrm{c}}' X^{(0)} - X^{(0)} (\tilde{H}_{\mathrm{v}}'-\hat{\Delta}) &=
  - [\tilde{H}_{\mathrm{c}}',k_{j}]X^{(1)}_{j} - [[\tilde{H}_{\mathrm{c}}',k_{j}] X^{(2)}_{jl} ,k_{l}]]. \label{Sylv-c}
\end{align}
\end{subequations}
The non-commutativity problem persists since $\tilde{H}_{\mathrm{c}}' $ and
$\tilde{H}_{\mathrm{v}}' $ contain $k$-dependent terms, while $X^{(n)}$ are
position-dependent.  Returning to Eq.~\eqref{S-kom} one can see that the terms of
$\tilde{H}^{(\mathrm{d})}$ linear and quadratic
in $k$ generate corrections to $S$ on the order of $k^{2}$ and $k^{3}$,
respectively. According to Eq.~\eqref{Ht}, these corrections generate terms
$O(k^{3})$ and $O(k^{4})$ in the effective mass Hamiltonian for the cb. The latter
are beyond the usual effective mass approximation, while the former correspond to
spin-orbit terms but (by a simple perturbation argument) appear with a coefficient 
$\sim P^{2} C_{k}/\Eg^{2}\sim 0.6$~nm$^{3}\cdot$meV (using InAs parameters), which is two
orders of magnitude smaller than the Dresselhaus coefficient for InAs,
$\alpha_{\mathrm{D}}=27$~nm$^{3}\cdot$meV. It appears, therefore, that the kinetic part of
$\tilde{H}_{\mathrm{v}}'$ can be discarded in the derivation of a usual effective mass equation in
the parabolic band approximation, that is, one
with $O(k^{2})$ kinetic terms and the relevant $O(k^{3})$ spin-orbit corrections. However,
as we will see below, including non-parabolicity effects at this point improves accuracy
of the modeling of a self-assembled QD. Therefore, it seems reasonable to keep the 
kinetic part of $\tilde{H}_{\mathrm{c}}'$ and 
$\tilde{H}_{\mathrm{v}}'$.  In order to obtain a solvable system of equations, we therefore
propose to self-consistently replace the $\kk$-dependent terms by their averages in the
eigenstate of interest.

In our derivation, 
Eqs.~\eqref{Sylv-a}--\eqref{Sylv-c} depend on the blocks of the transformed Hamiltonian
$\tilde{H}$ rather than on the initial Hamiltonian $H$ hence, together with
Eq.~\eqref{Ht}, they form a system that cannot be solved in a closed form. Clearly, in
the leading order one could replace $\tilde{H}_{\mathrm{c}}'$ and
$\tilde{H}_{\mathrm{v}}'$ by the original blocks $H_{\mathrm{c}}'$ and
$H_{\mathrm{v}}'$ (the correction is $O(k^{2})$, yielding corrections that are formally $O(k^{4})$ in the
resulting cb Hamiltonian). However, as we will see, including at least some
corrections to these blocks improves the accuracy of the equation. It is known that the
major corrections to the cb and vb Hamiltonians resulting from the
decoupling procedure are the renormalization of the cb electron mass and of the Luttinger
parameters, respectively. Therefore, we propose to take into account these strong effects
only and to use, in place of $\tilde{H}_{\mathrm{c}}'$ and $\tilde{H}_{\mathrm{v}}'$, the
cb and vb blocks of the original 8-band Hamiltonian but with the
position-dependent renormalized parameters,
\begin{equation}
 \frac{m_{0}}{\tilde{m}} =  \frac{m_{0}}{m'} + \frac{2}{3} \frac{E_{P}}{E_{g}'} +
                \frac{1}{3} \frac{E_{P}}{E_{g}'+\Delta_{0}}, \quad
  \tilde{\gamma}_{1} =  \gamma'_{1} + \frac{E_{P}}{3E_{g}'+\Delta_{0}},   \quad
  \tilde{\gamma}_{2,3} =  \gamma'_{2,3} + \frac{1}{2} \frac{E_{P}}{3E_{g}'+\Delta_{0}}, 
\label{renorm}
\end{equation}
where $E_{\mathrm{g}}'=E_{0}+(a_{\mathrm{c}}-a_{\mathrm{v}})\tr\eta$ is the local band
gap, including the hydrostatic strain-induced shift. 
Note that these band-decoupling corrections are $O(k^{2})$, so the distinction between
$\tilde{H}_{\mathrm{c,v}}'$ and 
$H_{\mathrm{c,v}}'$ is only important if the $k$-dependent terms are included
self-consistently, as proposed above.

In the approximations proposed here, Eqs.~\eqref{Sylv-a}--\eqref{Sylv-c} are a system of
usual Sylvester equations that can be solved iteratively. A solution can 
be obtained in a closed, analytical form using the general method of
Ref.~\cite{Hu2006}. However, the form of the solution simplifies considerably if one
discards the contribution of the spin-dependent part of $\tilde{H}_{\mathrm{c}}'$ to the operator
$S$ (these terms remain included to the leading order in $H_{\mathrm{6c6c}}$ that is part
of $H^{(\mathrm{d})}$, see
Eq.~\eqref{6c6c} and Eq.~\eqref{Ht}). These
terms are very small compared to any other energy scales in the problem, hence their
contribution to $S$ is negligible. Within this approximation, one has
$\tilde{H}_{\mathrm{c}}'=\chi_{\mathrm{c}}'\mathbb{I}_{2\times 2}$, where $\chi_{\mathrm{c}}'$ is a
scalar function of  position and $\mathbb{I}_{2\times 2}$ is a $2\times 2$ unit
matrix. Then, the solution to the above system of equations can be obtained
trivially. Denoting $\mathcal{D}=\hat{\Delta} + \chi_{\mathrm{c}}'\mathbb{I}_{6\times 6}
  -\tilde{H}_{\mathrm{v}}'$ one has
\begin{equation*}
X^{(2)}_{jl}  =   C^{(2)}_{jl} \mathcal{D}^{-1},   \quad
X^{(1)}_{j} =
  C^{(1)}_{j}\mathcal{D}^{-1} 
  +C_{jl}^{(2)} \left[k_{l},\mathcal{D}^{-1}\right],\quad
X^{(0)}  = C^{(1)}_{j}[k_{j},\chi_{\mathrm{c}}']\mathcal{D}^{-2}
-\left[k_{l},[k_{j},\chi_{\mathrm{c}}']C_{jl}^{(2)}\right]\mathcal{D}^{-2}.
\end{equation*}

From Eqs.~\eqref{Ht}, with the condition $\tilde{H}^{\mathrm{od}}=0$, one finds \cite{Shavitt1980}
$\tilde{H}^{\mathrm{d}}=H^{\mathrm{d}}+\tanh(\hat{S}/2) H^{\mathrm{od}}$ or, in the
leading order, $\tilde{H}^{\mathrm{d}}\approx H^{\mathrm{d}}+(1/2) [H^{\mathrm{od}},S]$.
Then, the correction to the cb Hamiltonian up to the order $k^{3}$ can be
decomposed into two parts. The 
first one, which we will denote by $\tilde{H}^{(2)}$, is formally quadratic in $\kk$ and is
proportional to $C_{j}^{(1)}C_{l}^{(1)}$ in our notation. This contribution yields
corrections to the electron effective mass and Land\'e factor. The second part, denoted
$\tilde{H}^{(3)}$, is of third 
order in $\kk$ and contains terms proportional to $C_{jl}^{(2)}C_{n}^{(1)}$. It
includes the Dresselhaus spin-orbit term. Thus, the resulting transformed
Hamiltonian can be written as 
$\tilde{H}=
H_{\mathrm{c}}+\tilde{H}^{(2)}+\tilde{H}^{(3)}$.
From Eq.~\eqref{Ht} in the linear approximation one finds
\begin{equation}
  \label{H2}
\tilde{H}^{(2)}  = \sum_{jl}k_{j}C^{(1)}_{j}\mathcal{D}^{-1}C^{(1)\dag}_{l}k_{l}
+\frac{1}{2}\sum_{jl}\left( 
[k_{j},\chi_{\mathrm{c}}']C^{(1)}_{j}\mathcal{D}^{-2}C^{(1)\dag}_{l}  k_{l} +\hc\right)
\end{equation}
and
\begin{equation}
  \label{H3}
\tilde{H}^{(3)}  = \sum_{jln}k_{j}C^{(2)}_{jl}k_{l}\mathcal{D}^{-1}C^{(1)\dag}_{n}k_{l} 
+\frac{1}{2}\sum_{jln}\left\{
-\left[k_{l},[k_{j},\chi_{\mathrm{c}}'] C^{(2)}_{jl}
  \right]\mathcal{D}^{-2}C^{(1)\dag}_{n}k_{n} 
+ k_{l} C^{(2)}_{jl} k_{j}\mathcal{D}^{-2}C^{(1)\dag}_{n}[k_{n},\chi_{\mathrm{c}}'] + \hc
\right\}. 
\end{equation}
The first term in Eq.~\eqref{H3} leads to the usual Dresselhaus spin-orbit coupling. The
other terms are linear or quadratic in $k$ and yield a small correction to the kinetic and
Zeeman terms of the effective mass Hamiltonian.

\section*{Interpretation, verification and discussion}

In this Section, we first present various approximations to the effective mass
Hamiltonian and relate the equations resulting from some
of these approximations to the common form of the effective mass Hamiltonian, written in
terms of the effective mass  
tensor and the $g$-factor given by the Roth formula.
Next we define a series of approximation for which we present
quantitative comparison of the predictions from the effective mass theory with the results
from the 8-band \kp Hamiltonian.

\subsection*{Interpretation of the effective mass equation}

 We restrict the discussion to the
quadratic term [Eq.~\eqref{H2}] that determines the fundamental properties of the
energy spectrum.  

In order to simplify notation, we define the $2\times 6$ matrices
$\mathcal{T}_{i}=\sqrt{3}T_{i}\oplus(-1/\sqrt{3})\sigma_{i}$, where the two components of
the direct sum correspond to the $j=3/2$ (hh and lh) and $j=1/2$ (spin-orbit split-off)
subbands of the vb. Then, by direct inspection
of Eq.~\eqref{6c8v} and Eq.~\eqref{6c7v} one finds $C_{j}^{(1)}=-P\ttcal_{j}$, where 
$\ttcal_{j}=\sum_{n}(\delta_{jn}-\eta_{jn})\tcal_{n}$.
The second-order correction to the effective mass Hamiltonian is then
\begin{equation}
  \label{H2-T}
\tilde{H}^{(2)}  = \sum_{jl}k_{j}P\ttcal_{j}\mathcal{D}^{-1}\ttcal^{\dag}_{l} P k_{l} +\frac{1}{2}\sum_{jl}\left(
P\ttcal_{j}\mathcal{D}^{-1}[k_{j},\chi_{\mathrm{c}}']\mathcal{D}^{-1}
\ttcal^{\dag}_{l}P  k_{l} +\hc\right),
\end{equation}
where we use the fact that 
$[k_{j},\chi_{\mathrm{c}}']$ is a number and commutes with the matrices $\ttcal_{i}$ and
$\mathcal{D}$. 

It might seem that the above Hamiltonian confirms the correctness of the particular
(BenDaniel-Duke\cite{BenDaniel1966}) 
ordering  of the operators in the kinetic term. This is not the case: a simple manipulation of
the terms in Eq.~\eqref{H2-T} under assumption $P=\mathrm{const}$ allows one to rewrite
the second-order correction in the equivalent form
\begin{equation}
  \label{H2-T-alt}
\tilde{H}^{(2)}  = 
\frac{1}{2}\sum_{jl}\left\{k_{j}k_{l},P\ttcal_{j}\mathcal{D}^{-1}\ttcal^{\dag}_{l} P
\right\} +\frac{1}{2}\sum_{jl}\left( 
P\ttcal_{j}\mathcal{D}^{-1}[k_{j},H_{\mathrm{v}}']\mathcal{D}^{-1}
\ttcal^{\dag}_{l}P  k_{l} +\hc\right),
\end{equation}
where a different (Gora-Williams-Bastard\cite{Gora1969,Bastard1975,Bastard1981}) ordering
appears in the kinetic term. 
Interestingly, the correcting terms in Eq.~\eqref{H2-T} and Eq.~\eqref{H2-T-alt} involve
only spatial derivatives of cb and vb parameters, respectively.
It is hence clear that the two
orderings are, in a sense, dual and each of them corresponds to neglecting terms in
the Hamiltonian, related to spatial modulation of either the conduction or valence
band. As we will see explicitly below, these additional terms are essential to correctly
reproduce the Rashba spin-orbit
interaction.

Further analytical insight into the somewhat unusual form of our effective mass
Hamiltonian $\tilde{H}$ is hindered by the need to invert the ``band gap operator''
$\mathcal{D}$, which produces rather intransparent and intractable formulas (and, in
practice, is performed numerically). One might
expect that neglecting the off-diagonal elements of $\mathcal{D}$, from which this
difficulty stems, is a good approximation, since these elements are rather small compared
to the band gap. The operator $\mathcal{D}$ in this approximation will be denoted by
$\tilde{\mathcal{D}}$. Since it is diagonal its inverse powers are found
trivially. Let us denote its diagonal elements by 
$(E_{\mathrm{hh}},E_{\mathrm{lh}},E_{\mathrm{lh}},E_{\mathrm{hh}},E_{\mathrm{so}},E_{\mathrm{so}})$. They
can be interpreted as the offset of the local edges of the three valence subbands with
respect to the cb edge at a given point in space.
We split the $\tcal_{j} \tilde{\mathcal{D}}^{-n}
\tcal_{l}^{\dag}$ matrix into a symmetric and asymmetric part, 
\begin{equation}\label{s-as}
\left(\ttcal_{j} \tilde{\mathcal{D}}^{-n} \ttcal_{l}^{\dag} \right)_{\mathrm{s},\mathrm{as}}  =
\frac{1}{2} \left(   \ttcal_{j} \tilde{\mathcal{D}}^{-n} \ttcal_{l}^{\dag} 
\pm  \ttcal_{l} \tilde{\mathcal{D}}^{-n}  \ttcal_{j}^{\dag}  \right).
\end{equation} 

Using the explicit forms of the matrices $\tcal_{j}$ one finds up to the linear order in
$\eta$ 
\begin{displaymath}
\left(\ttcal_{j} \tilde{\mathcal{D}}^{-n} \ttcal_{l}^{\dag} \right)_{\mathrm{s}}  =
\frac{\hbar^{2}}{2m_{0}P^{2}}
\left(\delta_{jl}-2\eta_{jl} \right)
\frac{F_{j}^{(n)}+F_{l}^{(n)}}{2}\mathbb{I}_{2\times 2},
\end{displaymath}
where $F_{x}^{(n)}=F_{y}^{(n)}=E_{\mathrm{P}}(2/E^{n}_{\mathrm{so}}+1/E^{n}_{\mathrm{lh}}
+3E^{n}_{\mathrm{hh}})/6$, 
$F_{z}^{(n)}=E_{\mathrm{P}}(1/E^{n}_{\mathrm{so}}+2/E^{n}_{\mathrm{lh}})/3$. 
Thus, the symmetric part is spin-independent and contributes to the
kinetic part of the effective mass Hamiltonian in the
well-known way via the electron effective mass tensor; indeed, the first term in
Eq.~\eqref{H2-T} combined with the kinetic term in $H_{\mathrm{6c6c}}$
[Eq.~\eqref{6c6c}] is 
\begin{equation}   \label{Hc-kinet1}
\tilde{H}^{(\mathrm{s},1)}  = \sum_{j}\frac{\hbar^{2}k_{j}^{2}}{2m_{j}^{*}},
\quad\mbox{where}\quad
\frac{m_{0}}{m^{*}_{j}}=\frac{m_{0}}{m'}+F_{j}^{(1)}.
\end{equation}
The second term yields
\begin{equation}   \label{Hc-kinet2}
\tilde{H}^{(\mathrm{s},2)} = 
-\frac{\hbar^{2}}{4m_{0}}\sum_{jl} \left[
k_{l},[k_{j},\chi_{\mathrm{c}}']
\left(\frac{1}{2}\delta_{jl}-\eta_{jl}\right)
\left(F_{j}^{(2)}+F_{l}^{(2)}\right)
\right]. 
\end{equation}
Thus, the symmetric component of the first term in Eq.~\eqref{H2-T} yields the usual kinetic
energy of 
the effective mass theory (Eq.~\eqref{Hc-kinet1}) in the commonly used ordering
``$k(m^{*})^{-1}k$'' (BenDaniel-Duke\cite{BenDaniel1966} ordering). This is, however,
corrected by 
the second term  (Eq.~\eqref{Hc-kinet2}) and one could as well start from
Eq.~\eqref{H2-T-alt} and arrive at an equivalent Hamiltonian with
Gora-Williams-Bastard\cite{Gora1969,Bastard1975,Bastard1981} ordering and with an
$H_{\mathrm{v}}$-dependent correction term instead of Eq.~\eqref{Hc-kinet2}.

The asymmetric part of $H^{(2)}$ can be interpreted most easily starting form the
alternative form of the Hamiltonian given in Eq.~\eqref{H2-T-alt}. 
The asymmetric contribution to the first term in Eq.~\eqref{H2-T-alt}, combined
with the Zeeman term of $H_{\mathrm{6c6c}}$ [Eq.~\eqref{6c6c}] can be
written in the form 
\begin{equation*}
\tilde{H}^{(\mathrm{as},1)}= 
\frac{1}{4}\sum_{jl} \left\{ k_{j},\left[k_{l} , \mathcal{A}_{jl}\right] \right\},
\quad\mbox{where}\quad
\mathcal{A}_{jl} = 
\frac{i\hbar^{2}g'}{4m_{0}} \sum_{m}\varepsilon_{jlm} \sigma_{m}+
P\left(\ttcal_{j}\mathcal{D}^{-1}\ttcal^{\dag}_{l}\right)_{\mathrm{as}} P.
\end{equation*}
In the diagonal approximation, $ \mathcal{D}\approx  \tilde{\mathcal{D}}$, the asymmetric
part yields up to order $O(\eta)$
\begin{equation}\label{TDT-as}
\left(\ttcal_{j} \tilde{\mathcal{D}}^{-n} \ttcal_{l}^{\dag} \right)_{\mathrm{as}}  =
-\frac{i}{2E_{\mathrm{P}}}\sum_{m}
\left[
\varepsilon_{jlm} - \sum_{n}(\eta_{jn}\varepsilon_{nlm} +\eta_{nl}\varepsilon_{jnm} )
\right]
\delta g_{m}^{(n)}\sigma_{m}, 
\end{equation}
where 
\begin{equation}\label{gxyz}
\delta g_{x}^{(n)}=\delta
  g_{y}^{(n)}  =(2/3)E_{\mathrm{P}}(1/E_{\mathrm{lh}}^{n}-1/E_{\mathrm{so}}^{n}), 
\delta g_{z}^{(n)} =E_{\mathrm{P}}[-1/(3E_{\mathrm{lh}}^{n})+1/E_{\mathrm{hh}}^{n} -
2/(3E_{\mathrm{so}}^{n})].
\end{equation} 
Using the relation 
$\left[k_{j},k_{l}\right]=-\frac{ie}{\hbar}\varepsilon_{jlm}B_{m'}$,
and defining the (diagonal) Land\'e tensor 
$\hat{g}_{mm'}=g'\delta_{mm'}
-\left[ \delta_{mm'}(1+\tr\eta)-\eta_{mm'}\right]\delta g_{m}^{(1)}$, 
one gets the usual Zeeman Hamiltonian
\begin{equation}\label{Has1}
\tilde{H}^{(\mathrm{as},1)}  \approx  \frac{1}{2}\mu_{\mathrm{B}}
                            \bm{B}\hat{g}\bm{\sigma}.
\end{equation}
The form of the Land\'e tensor $\hat{g}$ appearing in this equation is
close to the widely used isotropic Roth-Lax-Zwerdling\cite{Roth1959} formula 
\begin{equation}\label{Roth}
g^{(\mathrm{Roth})}=g'- \frac{2E_{\mathrm{P}}}{3}\left(\frac{1}{\Eg}-\frac{1}{\Eso}\right).
\end{equation}
Here, $\hat{g}$ differs from $g^{^{(\mathrm{Roth})}}$ by including strain effects as well as
anisotropy resulting from the hh-lh splitting in the nanostructure. 
The Roth-Lax-Zwerdling formula
has the advantage that the fundamental band 
gap $\Eg$, instead of being calculated at some level of approximation to the \kp theory,
can be taken from experiment\cite{syperek11,Syperek2012,VanBree2012}. 
While the standard Roth-Lax-Zwerdling formula is linked to the 8-band \kp
Hamiltonian via a diagonal and parabolic approximations, the full effective mass
Hamiltonian accounts also for non-parabolicity corrections and the full structure of the vb.

The asymmetric part of the second term in Eq.~\eqref{H2-T-alt} is
a generalized Rashba term. To see this, assume that the spatial variation of
$H_{\mathrm{v}}'$ results from an external electric field $\bm{\mathcal{E}}$, hence 
$[k_{j}, H_{\mathrm{v}}']=-i\nabla H_{\mathrm{v}}'=-ie\bm{\mathcal{E}}\mathbb{I}_{6\times 6}$. 
Then, taking only the strain-independent terms in the second term in Eq.~\eqref{H2-T-alt}
and using Eq.~\eqref{TDT-as} one 
finds in the diagonal approximation
\begin{displaymath}
H^{(\mathrm{as,2})} = \frac{eP^{2}}{4E_{\mathrm{P}}}\sum_{jlm}\mathcal{E}_{j}
\varepsilon_{jlm}\delta g_{m}^{(2)}\sigma_{\mathrm{m}}k_{l}+\hc
\end{displaymath}
Neglecting the anisotropy due to the lh-hh splitting in Eq.~\eqref{gxyz} one finds
\begin{displaymath}
H^{(\mathrm{as,2})} = \frac{eP^{2}}{3}
\left( \frac{1}{E_{\mathrm{g}}^{2}} - \frac{1}{(E_{\mathrm{g}}+\Delta_{\mathrm{SO}})^{2}}
\right)
\frac{1}{2}\bm{\mathcal{E}}\cdot (\bm{\sigma}\times\bm{k}) + \hc,
\end{displaymath}
which is the standard Rashba spin-orbit term \cite{winkler03}. 
The full Eq.~\eqref{H2-T-alt} yields the Rashba term generalized to
arbitrary inhomogeneity of the valence band and includes corrections due to inhomogeneous
strain.  The same effects are accounted for (although not so explicitly) by Eq.~\eqref{H2-T}.

\subsection*{Quantitative assessment}

In the following, we compare the accuracy (with respect to the 8-band \kp results) of
various approximations to the effective mass equation derived above. Our systematic
approach allows us to discuss
the possible approximate equations in the uniform framework, according to the level of
approximation made to the $\mathcal{D}$ operator.

{\parindent0mm
(1) As the very first step, obviously inaccurate but included for completeness, we apply the
\textit{bulk approximation without strain}, in which the operator $\mathcal{D}$ is
taken diagonal and without the kinetic ($k^{2}$) terms,
\begin{equation}\label{D-diag}
\mathcal{D} = \tilde{\mathcal{D}}=
\mathrm{diag}(E_{\mathrm{hh}},E_{\mathrm{lh}},E_{\mathrm{lh}},
E_{\mathrm{hh}},E_{\mathrm{so}},E_{\mathrm{so}}),
\end{equation}
with local bad gaps calculated from unstrained bulk values, 
$E_{\mathrm{hh}}=E_{\mathrm{lh}}=E_{0}$, $E_{\mathrm{so}}=E_{0}+\Delta_{0}$,
where the band edges are interpolated
according to the local composition (see Tab.~\ref{tab:param}). 

(2) The first reasonable approximation is to correct the above procedure by accounting for
band-edge shifts due to local strain, which is done by including the diagonal
strain-dependent matrix elements in $H_{\mathrm{v}}$ and $H_{\mathrm{c}}$. Thus, the
$\mathcal{D}$ operator is still given by Eq.~\eqref{D-diag} but with
$E_{\mathrm{hh}}=E_{0}+(a_{\mathrm{c}}-a_{\mathrm{v}})\tr\eta
+b_{\mathrm{v}}(2\eta_{zz}-\eta_{xx}-\eta_{yy})/2$, 
$E_{\mathrm{lh}}=E_{0}+(a_{\mathrm{c}}-a_{\mathrm{v}})\tr\eta
-b_{\mathrm{v}}(2\eta_{zz}-\eta_{xx}-\eta_{yy})/2$, 
$E_{\mathrm{so}}=E_{0}+\Delta_{0}+(a_{\mathrm{c}}-a_{\mathrm{v}}) \tr\eta$,
where the local value of strain is used and parameters are interpolated according to the
local composition.
In this way, the major correction to the band gap (which is a
crucial factor determining the effective mass and $g$ factor) is included in the
model. This will be referred to as \textit{bulk approximation with strain}.

(3) The value used for the band gap in the previous approach is still a crude
approximation since the actual energy spacing between the states in the conduction and
valence bands in a QD are affected by spatial confinement. A common approach,
taken in many cases\cite{syperek11,Syperek2012,VanBree2012} to estimate the electron
$g$-factor when interpreting experimental data, is to 
use the measured  value of the fundamental transition energy $\Eg^{(\mathrm{exp})}$ as the
effective band gap value for a given system. In our numerical study this 
corresponds to replacing the fundamental  band edge offset in the operator $\mathcal{D}$
(together with its strain correction) by the splitting between the top vb and bottom cb
states obtained from the full 8-band calculation, while using the bulk values of the
spin-orbit splitting $\Delta_{0}$ and the Kane parameter $P$ interpolated according to
composition. The $\mathcal{D}$ operator is still given by the diagonal form of
Eq.~\eqref{D-diag} but now
$E_{\mathrm{hh}}=E_{\mathrm{lh}}=\Eg^{(\mathrm{exp})}$, 
$E_{\mathrm{so}}=\Eg^{(\mathrm{exp})}+\Delta_{0}$,
with a constant $\Eg^{(\mathrm{exp})}$ and a local composition-dependent $\Delta_{0}$.
We will refer to this as \textit{semi-phenomenological approximation}.  
For the $g$-factor, a further simplification along this line is achieved by
introducing constant values of $\Delta_{0}$ obtained by averaging the
position-dependent values weighted by the squared wave function, which yields an
explicit \textit{effective Roth-Lax-Zwerdling formula} for the $g$-factor of a
nanostructure, that is, Eq.~\eqref{Roth} with the effective values of the parameters.

(4) In the \textit{off-diagonal} approximation, we include the full structure of the
vb Hamiltonian $H_{\mathrm{v}}$, hence the full form of $\mathcal{D}$, but still
neglect the $k$-dependent terms. In block notation,
\begin{equation*}
\mathcal{D}=(E_{\mathrm{c}}+V_{\mathrm{p}}+a_{\mathrm{c}}\tr\eta)\mathbb{I}-
\left(\begin{array}{cc}
H_{\mathrm{8v8v}}^{(x)} & H_{\mathrm{8v7v}}^{(x)} \\
H_{\mathrm{7v8v}}^{(x)} & H_{\mathrm{7v7v}}^{(x)} \\
\end{array}\right),
\end{equation*}
where $H_{\mathrm{8v8v}}^{(x)}$, $H_{\mathrm{8v7v}}^{(x)}$, and $H_{\mathrm{7v7v}}^{(x)}$
are the position-dependent 
parts of the corresponding vb Hamiltonian blocks, formally obtained by setting $k=0$ in
Eq.~\eqref{8v8v}, Eq.~\eqref{8v7v}, and Eq.~\eqref{7v7v}, respectively, and
$H_{\mathrm{7v8v}}^{(x)} = H_{\mathrm{8v7v}}^{(x)\dagger}$. 
This yields a cb Hamiltonian
strictly quadratic in $k$ (and strictly equivalent to the original 8-band Hamiltonian up
to the quadratic order),
which corresponds to the usual notion of the effective mass 
equation in the parabolic approximation. Starting from this approximation, we go beyond
the usual forms of the effective mass equations by introducing the full matrix structure
of the vb-induced corrections.

(5) The next step consists in including the self-consistent averages of the $k$-dependent
terms in $\mathcal{D}$, that is
\begin{equation}\label{D-off-diag}
\mathcal{D}  =\left(E_{\mathrm{c}}+V_{\mathrm{p}}+a_{\mathrm{c}}\tr\eta 
+\frac{\hbar^{2}}{2m'} \left\langle k_{x}k_{x} +\cp\right\rangle \right)\mathbb{I} 
- \left(\begin{array}{cc}
H_{\mathrm{8v8v}}^{(\mathrm{av})} & H_{\mathrm{8v7v}}^{(\mathrm{av})}  \\
H_{\mathrm{7v8v}}^{(\mathrm{av})} & H_{\mathrm{7v7v}}^{(\mathrm{av})} \\
\end{array}\right),
\end{equation}
where $H_{\mathrm{8v8v}}^{(\mathrm{av})}$, $H_{\mathrm{8v7v}}^{(\mathrm{av})}$, and
$H_{\mathrm{7v7v}}^{(\mathrm{av})}$ are obtained from 
Eq.~\eqref{8v8v}, Eq.~\eqref{8v7v}, and Eq.~\eqref{7v7v}, respectively, by
self-consistently replacing 
each product $k_{i}k_{j}$ with its average (denoted by $\langle\ldots\rangle$) in
the state of interest. In this way we go beyond the parabolic 
approximation. We will call this the \textit{off-diagonal}$+\langle k^{2}\rangle$
approximation. In this approximation the implicit character of
Eqs.~\eqref{Sylv-a}--\eqref{Sylv-c} (the appearance of $\tilde{H}_{\mathrm{c}}$ and
$\tilde{H}_{\mathrm{v}}$) is resolved to the leading order, that is, by replacing the two
blocks of the Hamiltonian by the original $H_{\mathrm{c}}$ and $H_{\mathrm{v}}$. 

(6) The final approach in our sequence of approximations consists in approximating
$\tilde{H}_{\mathrm{c}}$ and $\tilde{H}_{\mathrm{v}}$ in
Eqs. ~\eqref{Sylv-a}--\eqref{Sylv-c} self-consistently by renormalizing the effective
mass and Luttinger parameters according to Eqs.~\eqref{renorm}. The form of $\mathcal{D}$
is therefore the same as in 
Eq.~\eqref{D-off-diag} but with the renormalized parameters. We will refer to this as the
\textit{self-consistent effective mass equation}. 
}

\begin{figure*}
	\begin{center}
		\includegraphics[width=.8\textwidth]{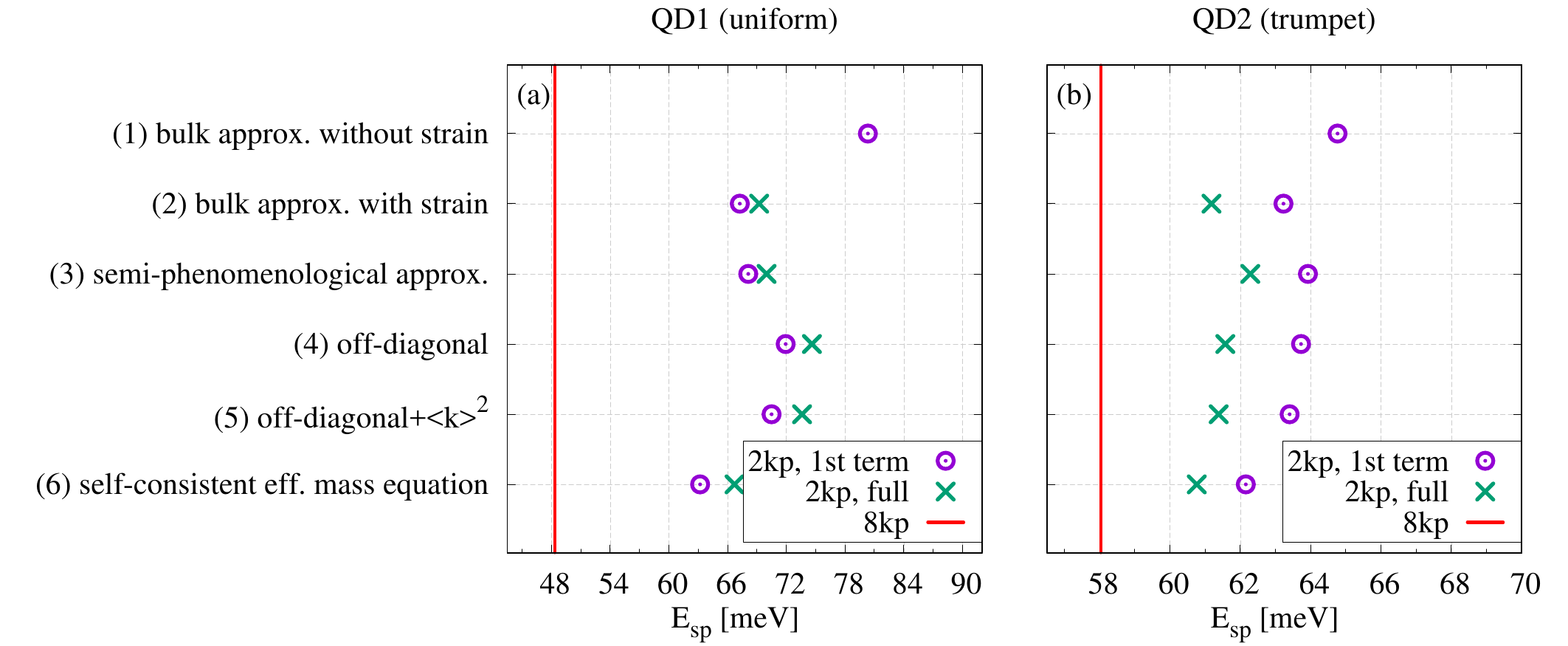}
	\end{center}
	\caption{\label{fig:Esp}Energy splitting between the ground and first excited
          states at zero magnetic field for the sequence of approximations. Dots show the
          results obtained using only the first term in Eq.~\eqref{H2-T}, while crosses
          represent the results from the full Hamiltonian. The red line shows the value
          obtained from the 8-band \kp calculation.}
\end{figure*}

\begin{figure*}
	\begin{center}
		\includegraphics[width=.8\textwidth]{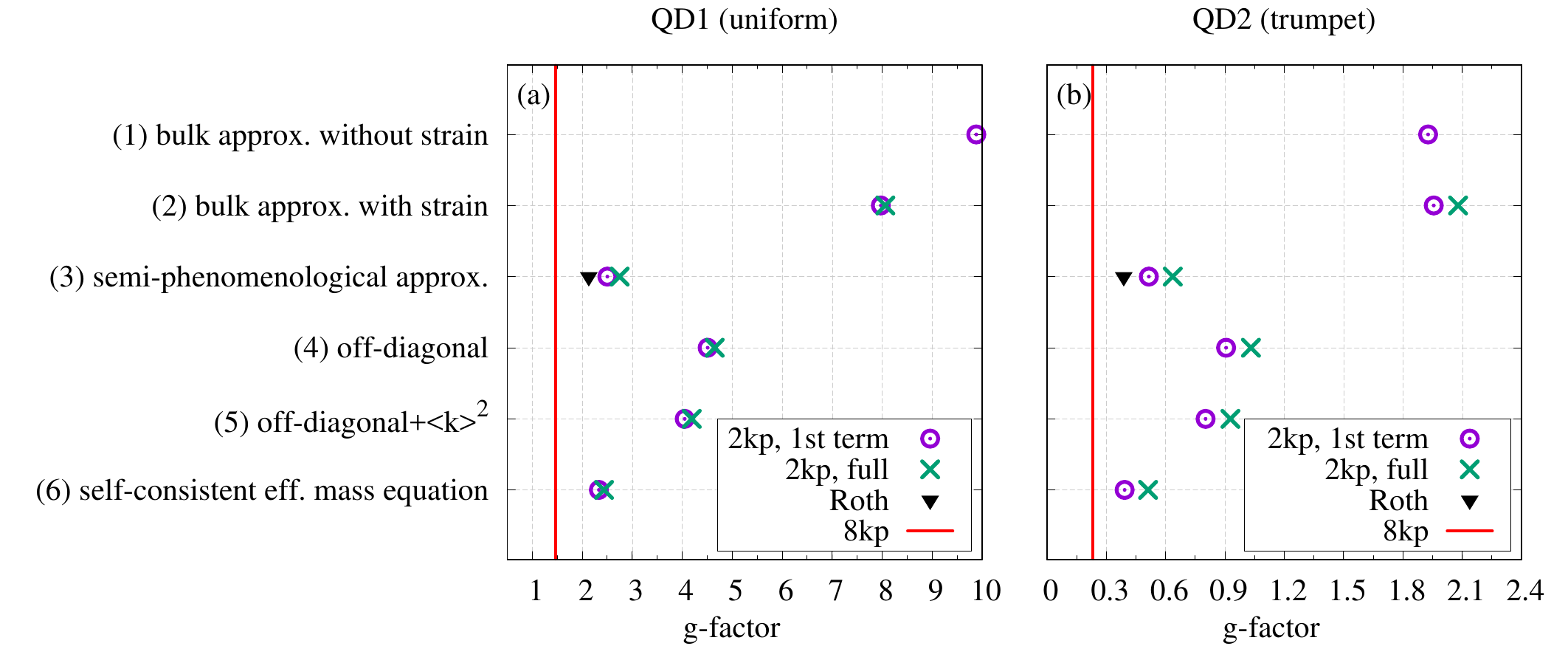}
	\end{center}
	\caption{\label{fig:g}Ground state Land\'e factor for the sequence of
          approximations. Dots and crosses are defined as in Fig.~\ref{fig:Esp}. The triangle
          shows the result from the effective Roth formula. The red line shows the value
          obtained from the 8-band \kp calculation.}
\end{figure*}

As a test of these approximations, we study the low-energy part of the
spectrum of a self-assembled QD. 
We consider two models of a lens-shaped, self-assembled InAs/GaAs QD. In both models the
QD has 
$24$~nm diameter and $4.2$~nm height but they differ in the composition profile: While the
first model assumes a uniform composition with 100\% InAs inside the QD, the second one
has a more realistic trumpet-shape profile of InAs/GaAs composition with the InAs content
defined by \cite{migliorato02} 
$C(\rr) = C_{\mathrm{b}} + (C_{\mathrm{t}}-C_{\mathrm{b}}) \exp[ 
    -\sqrt{x^{2}+y^{2}} \exp(-z/z_{\mathrm{p}})/r_{\mathrm{p}}]$,
where we took $C_{\mathrm{b}}=0.4$ , $C_{\mathrm{t}}=0.8$, $r_{\mathrm{p}}=0.9$~nm and
$z_{\mathrm{p}}=1.4$~nm. 
In both
cases   the QD is placed on a $0.6$~nm 
wetting layer (WL), which in the case of the first dot contains 100\% and in the latter
one 40\% InAs.  The two figures of merit that we will investigate here are the energy 
splitting $\Delta E_{sp}$ between the ground (``$s$-shell'') electron state
and the lowest excited (``$p$-shell'') state at 
zero magnetic field (Fig.~\ref{fig:Esp}) and the ground state  $g$-factor, extracted from
the leading order 
(linear) term of the Zeeman splitting of the ground-state doublet at low magnetic fields
(Fig.~\ref{fig:g}). In both cases we show in the plots the values obtained using only the
first, usual term in Eq.~\eqref{H2-T} (points) and those from the full equation, including
also the non-standard second term (crosses).

As shown by the results in Fig.~\ref{fig:Esp}, the effective mass methods typically fail
to reproduce the $s$-$p$ shell splitting which, in fact, belongs to the most fundamental
quantitative characteristics of a QD system. What matters here is the
renormalization of the \kp parameters according to Eqs.~\eqref{renorm}, hence the
self-consistent equation (method (6) from our series 
of approximations) is able to produce a result that is closest to the correct value. For the
uniform QD,  Fig.~\ref{fig:Esp}(a), the disagreement is still at the level of 30\%, while
for the more realistic model with the trumpet-shape composition profile
Fig.~\ref{fig:Esp}(b), the effective mass 
result is much more exact. The 30\% discrepancy is consistent with the estimate of the
error induced by the linear truncation in Eq.~\eqref{S-kom}. The much higher precicision
in the case of the smooth trumpet-shape composition may be due to the fact that terms of
higher order in $k$ in an envelope-function theory for an inhomoheneous system are not just
``non-parabolic'' but also induce higher order derivatives of system parameters, to which
the uniform QD model with abrupt compositional boundaries is obviously much more
sensitive. One can understand why approximation (6) is particularly
suited for relatively correct modeling of this particular spectral feature by noting that
the $s$-$p$ shell splitting is related to the in-plane excitation of the QD, for which the
value of the effective mass plays a crucial role. The renormalization described by
Eq.~\eqref{renorm} is an important correction to this parameter. On the contrary,
including the non-parabolicity correction, either by self-consistently adding the average
$k^{2}$ terms in the $\mathcal{D}$ operator, as in the approach (5), or by using a
phenomenological value of $\Eg$, as in the approach (3), mostly reflects the impact of the strong
confinement in the growth direction, which shifts the lowest shells rigidly, hence does
not affect the $s$-$p$ splitting considerably. 
It is found
also that including the full matrix structure of $H_{\mathrm{v}}$ in $\mathcal{D}$
(approximation (4)) does not bring any clear advantage by itself. In addition, it turns out
that correcting the effective mass equation by the second term in Eq.~\eqref{H2-T} has an
opposite effect on the two models: it decreases the accuracy for the homogeneous QD, while
it improves it (to a larger degree) for the trumpet-shape model. We were not able to find
a plausible explanation of this fact. Interestingly, the correction stemming from this term
is roughly the same for a given QD structure in each approximation.

The accuracy of various approximations to the effective mass equation is different when
the $g$-factor is considered (Fig.~\ref{fig:g}). Here the approximations (4-6), where
the full matrix structure of the vb is included, yield much more accurate results than
the diagonal approximations (1) and (2). As one can see in
(Fig.~\ref{fig:g}), the self-consistent model (6), with self-consistently renormalized
parameters, again 
provides the most accurate results in the systematic series of approximations. The
discrepancy is about 60\% for the uniform QD model 
(Fig.~\ref{fig:g}a) while the result for the model with a trumpet-shape composition profile
(Fig.~\ref{fig:g}b) is overestimated by a factor of 2.  In the case of the $g$-factor,
including the part of
non-parabolicity that is captured by our self-consistent approach improves the accuracy to
some extent but is not critical. Also 
the correction provided by the second term in  Eq.~\eqref{H2-T} is of much less importance
than it was for the spectrum, with a larger relative contribution in the
trumpet-shape model. In both models, it turns out to lower the accuracy. 
The large error in the values of the $g$-factor apparently exceeds the estimate of the
effect of truncating  Eq.~\eqref{S-kom}. In fact, however, the numerical values should be
referred to the bulk value of $-15$, which is nearly entirely compensated by confinement and strain
effects. Therefore, the mismatch is actually of only a few percent.

It seems
interesting that accurate systematic modeling of the Zeeman 
splitting in an effective mass equation requires the full form of the operator
$\mathcal{D}$. This means that the standard Eq.~\eqref{Has1} is not very useful if the
$g$-factor is calculated by L\"owdin perturbation theory including only the diagonal
contributions to the vb structure (in particular, the hh-lh splitting induced by
the axial-strain-related $b_{\mathrm{v}}$ terms): for our two structures the approach (2)
overestimates the Zeeman splitting by an order of magnitude. 
 We were able to trace this back to the important role played by the
off-diagonal terms proportional to the $b_{\mathrm{v}}$ deformation
potential in the correct alignment of valence bands: including axial strain
contributions only in the diagonal terms produces wrong vb alignment (in terms of band
splitting and the sign of the band offsets between the QD and the barrier).
As a consequence, completely neglecting
these subtle details of the vb, as done in the semi-phenomenological
approximation (3), leads to a result which is only slightly worse than the most exact
one when compared to the full 8-band model. Here the lh-hh splitting and band offsets are
completely washed out by setting a 
single constant value for the fundamental band gap and only the spin-orbit ($\Gamma_{7}$)
band splitting is position dependent. The result becomes even more accurate if the latter
is also replaced by a single number obtained from spatial averaging of the standard bulk
values (black triangles in (Fig.~\ref{fig:g})). Thus, our numerical results support the
commonly used and very convenient 
Roth-Lax-Zwerdling formula for estimating the electron $g$-factor. 

\section*{Conclusions}

In this paper we have proposed a systematic derivation of a cb effective mass
equation from the 8-band \kp Hamiltonian. We have shown that this derivation develops
into a systematic series of approximations that differ in the way the vb is
represented in the final equation. Possible approximations range from using a set of fixed
band gap parameters to a full matrix structure with self-consistent non-parabolicity
corrections and parameter renormalization. We have assessed the accuracy of the
approximations in calculating selected spectral and spin-related
characteristics for a self-assembled QD system within two models of the composition
profile. 

We have shown that a quantitatively correct description of the lowest sector of the
electron spectrum, which involves intraband dynamics and therefore relies primarily on
the accurate modeling of the effective mass, requires a self-consistent renormalization of
the Hamiltonian parameters that goes beyond the second order L\"owdin perturbation. The
accuracy is also improved by accounting for cb non-parabolicity by
self-consistently including terms of higher order in the electron momentum.  When
studying the Zeeman splitting of the ground electron level we found that the most accurate
value of the $g$ factor is obtained within the systematic scheme only after including the
full structure 
of the vb Hamiltonian in the equation. Here, again, including non-parabolicity
corrections improves the accuracy of the result. Surprisingly, the values obtained from
the resulting rather complicated equation can be reproduced by a simple diagonal model
with a fixed 
value of the fundamental band gap, which supports the use of the effective
Roth-Lax-Zwerdling formula for a nanostructure. The accuracy of the latter is remarkable,
given its simplicity.  The effective mass equations reproduce 8-band \kp value 
of the electron $g$-factor for the QD ground state within a factor of 2, which may seem
disappointing. One should take into account, however, that the value of the $g$-factor for
bulk InAs is about $-15$, which is nearly entirely compensated by confinement and strain
effects. The absolute mismatch, which is on the order of 0.1, is a tiny fraction of the
original value (that is, the compensation is quantitatively reproduced) but yields a large
relative error when compared to the very small final value. 

In general, we have shown that the effective mass equation offers a limited accuracy when
modeling the lowest-energy sector of the electron spectrum in a self-assembled QD,
unless it is extended to a rather complicated, inconvenient and computationally expensive
form. In particular, an equation that is  rigorously derived from the  \kp theory strictly 
up to order $k^{2}$ (which may correspond to the most usual notion of an \textit{effective
  mass theory}) quantitatively fails in all respects. The more accurate
self-consistent equation proposed in this paper is not particularly transparent and does
not even allow a separation into kinetic and Pauli terms.
On the other
hand, we have shown that the electron $g$-factor can be estimated by a very
simple version of the effective mass theory, thus justifying the phenomenological
Roth-Lax-Zwerdling formula for a nanostructure. Our derivation highlights the
correspondence between the BenDaniel-Duke and Gora-Williams-Bastard orderings of the
kinetic term, both emerging as different approximations within a general scheme, and
yields a generalized Rashba coupling including inhomogeneity and strain effects. This
confirms once more that the effective mass equation, even though not perfectly accurate,
can be very useful from the conceptual point of view and often provides valuable physical
insight. 

\section*{Data availability}
The datasets generated during and/or analysed during the current study are available from
the corresponding author on reasonable request. 
   
\section*{Acknowledgments}
This work was supported by the
Polish National Science Centre (Grant No.~2014/13/B/ST3/04603). 
Calculations have been partly carried out in Wroclaw Centre for
Networking and Supercomputing (http://www.wcss.wroc.pl), Grant No. 203.
A.~M.-P. acknowledges support from the subsidy granted by the Polish MNiSW to the Faculty
of Fundamental Problems of Technology for the research activity contributing to the
development of young researchers and PhD students. 

\section*{Author contributions statement}
All authors performed analytical calculations. K.G. developed most of the code,
with contributions from A.M.-P. Simulations were performed by A.M.-P. and K.G. All authors
contributed equally to analyzing and interpreting the results. All authors contributed to
writing and reviewing the manuscript.

\section*{Additional information}

The authors declare that they have no competing interests.


\end{document}